\documentclass[prl,twocolumn,superscriptaddress]{revtex4-1}

\usepackage{graphicx}
\usepackage{dcolumn}
\usepackage{bm}
\usepackage{color}

\newcommand{\hto}[1]{Ho$_2$Ti$_2$O$_7${#1}}
\newcommand{\dto}[1]{Dy$_2$Ti$_2$O$_7${#1}}

\begin{document}


\title{Phonon mediated spin flipping mechanism in the spin ices \dto{} and \hto{}}
\author{M Ruminy}
\email{martin.ruminy@gmx.de}
\affiliation{Laboratory for Neutron Scattering and Imaging, Paul Scherrer Institut, 5232 Villigen PSI, Switzerland}
\author{S Chi}
\affiliation{Quantum Condensed Matter Division, Oak Ridge National Laboratory, Oak Ridge, Tennessee 37831, USA}
\author{S Calder}
\affiliation{Quantum Condensed Matter Division, Oak Ridge National Laboratory, Oak Ridge, Tennessee 37831, USA}
\author{T Fennell}
\email{tom.fennell@psi.ch}
\affiliation{Laboratory for Neutron Scattering and Imaging, Paul Scherrer Institut, 5232 Villigen PSI, Switzerland}

\date{\today}

\begin{abstract}
To understand emergent magnetic monopole dynamics in the spin ices \hto{} and \dto{}, it is necessary to investigate the mechanisms by which spins flip in these materials.  Presently there are thought to be two processes - quantum tunneling at low and intermediate temperatures, and thermally activated at high temperatures.  We identify possible couplings between crystal field and optical phonon excitations and construct a strictly constrained model of phonon-mediated spin flipping that quantitatively describes the high temperature processes in both compounds, as measured by quasielastic neutron scattering.   We support the model with direct experimental evidence of the coupling between crystal field states and optical phonons in \hto{}.  
\end{abstract}

\maketitle

In rare earth compounds, magnetic responses can be strongly and non-monotonically dependent on the strength or frequency of applied magnetic field, or the temperature.  Examples include stepped magnetization curves in single ion magnets~\cite{ishikawa:kx}, or the multiply-peaked susceptibility response in LiYF$_4$:Ho$^{3+}$~\cite{Giraud:2001cp,Giraud:2003il,Bertaina:2006gf}.  These effects appear because there are competing mechanisms that can contribute to the flipping of large rare earth magnetic moments.  Owing to their different origins - conduction electrons~\cite{Becker:1977ge}, phonon-mediated (e.g.  direct, Raman, Orbach, and phonon bottleneck effects~\cite{ORBACH:1961cd,Finn:2002fs,Scott:1962it}), or quantum mechanical (tunneling, thermally assisted tunneling between excited states, resonant tunneling at electronic-nuclear level crossings, and co-tunneling~\cite{Thomas:1996hg,Bokacheva:2000ey,Giraud:2001cp,Giraud:2003il,Gatteschi:bn}) - these mechanisms have quite different parametric dependencies.  Understanding spin flipping (or relaxation) is currently important in rare-earth based single ion magnets~\cite{Blagg:2013bf}, especially in the context of applications in quantum information processing~\cite{Leuenberger:2001ft,Ardavan:2007ci,Bogani:2008hi} that depend on the stability and control of quantum states~\cite{Prokofev:2000bd,feynmann,Ghosh:2002wr,Stamp:2004kf}, and in spin ices, where they determine the mobility of magnetic monopole excitations~\cite{Castelnovo:2008hb}.

In a canonical spin ice such as \dto{} or \hto{}~\cite{Bramwell:2001tpa}, the magnetization dynamics of the low temperature Coulomb phase~\cite{Fennell:2009ig,Henley:2010vo} should be described by the cooperative behavior of the thermal population of emergent magnetic monopoles~\cite{Castelnovo:2008hb,Ryzhkin:2005ko}, which form a magnetic Coulomb gas.  Indeed, the spin relaxation time, $\tau$, of \dto{}, as extracted from $\chi_{ac}$, has been explained with considerable success by the monopole picture: both a thermally activated regime at $T<1$ K~\cite{Snyder:2001uu,Matsuhira:2001cp,Snyder:2003ek,Snyder:2004it} (which we call low temperature) and temperature independent plateau for $1<T<10$ K (intermediate temperature) are captured well by a theory of monopole hopping in dilute (unscreened) and concentrated (screened) magnetic Coulomb gases respectively~\cite{Jaubert:2009ed}.  The reentrant low temperature thermal activation is due to interactions between unscreened monopoles. 

For a monopole to hop, a spin must be flipped, and because the plateau of $\tau$ was previously associated with quantum tunneling of the large, Ising-like Dy$^{3+}$ moments between the members of their ground state doublet~\cite{Snyder:2004it}, monopoles were assumed to hop by tunneling of the spins with temperature independent attempt frequency~\cite{Jaubert:2009ed}.  The resulting picture should describe the Coulomb gas realized in each material by relating the energy for monopole creation and unbinding to the exchange interactions~\cite{BrooksBartlett:2014kf,Zhou:2011fq}.  However, subsequent measurements of \hto{}~\cite{Quilliam:2011df} and \dto{}~\cite{Matsuhira:2011ci,Yaraskavitch:2012hk,Kassner:2015jo} have found that in the unscreened regime this relationship is not exactly as expected, while simulations of \dto{} with a temperature-dependent hop rate agree better with the observed relaxation times~\cite{Takatsu:2013bq}.  These studies suggest that to understand out-of-equilibrium~\cite{Giblin:2011ft,Revell:2012hq,Paulsen:2014cc,Paulsen:1970de} and quantum dynamics~\cite{Tomasello:2015vg,Rau:2015jj} in spin ices at low temperature, it is essential to understand all contributions to the monopole hopping dynamics.  As in LiYF$_4$:Ho$^{3+}$~\cite{Bertaina:2006gf}, the first requirement is to understand the classical spin flipping mechanism of the spin ices.

Studies of \dto{}~\cite{Snyder:2001uu,Matsuhira:2001cp,Snyder:2003ek,Snyder:2004it} in which the intermediate temperature plateau was ascribed to quantum tunneling of the spins also revealed a second thermally activated regime for $T>10$ K (i.e. high temperature).
The response of \hto{} is similar but the relative rate of the low temperature process is much faster than in \dto{}~\cite{Ehlers:2002jz,sup_mat}.   The high temperature process in both spin ices was modeled by an Arrenhius law, with activation energy $\Delta$ attributed to over-barrier hopping via the first crystal field excitation (CFE).  However, the best-fitting $\Delta$, although close, is not equal to the energy of any CFE in either material, and this interpretation does not explain how such a process would occur.  

\begin{figure}
  \centering
   \includegraphics[trim=1 1 1 1,clip=true, scale=1.0]{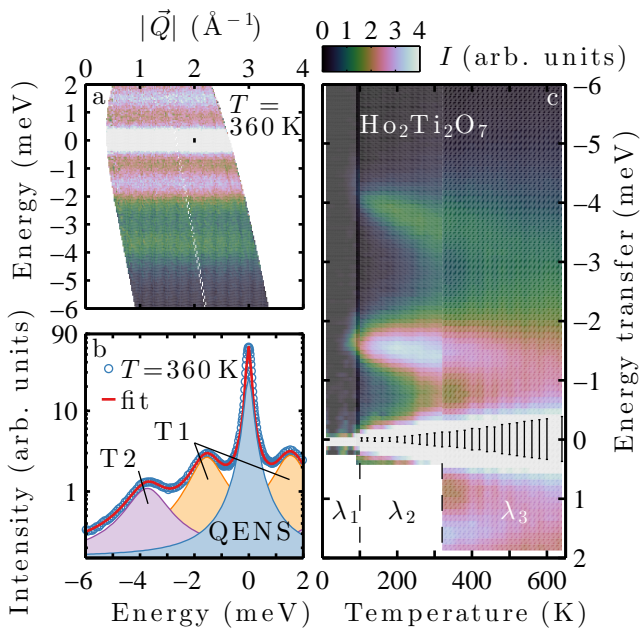}%
  \caption{General features of QENS in spin ices, as exemplified by \hto{}.  We see a $|\vec{Q}|$-independent $S(|\vec{Q}|,\omega)$ response, with QENS  around the elastic line and transitions amongst excited crystal field states (a).  In panel b, we show an example of a resolution-convoluted fit of the quasielastic Lorentzian (QENS) and two CFEs (T1 and T2).  In panel c we show the general evolution of the QENS and excited state transitions, along with the resolution regimes used in the measurements ($\lambda_{1,2,3}=11,6,4.3$ \AA ~in this case).}
   \label{fig:f1}
\end{figure}

We propose that  phonon mediated processes involving a higher crystal field state interacting with phonons~\cite{Lovesey:2000jh} provide a quantitative and  physical explanation of the high temperature processes.  In this mechanism, a rare earth ion is excited from one crystal field state to an intermediate excited state by absorption of a phonon, and then relaxes to a third state by emission of another phonon.    Relaxation by a single such process has the characteristic temperature dependence of $n=1/(\exp{(\Delta/k_BT)}-1)$, where $\Delta$ is the energy of the phonon to be absorbed, but more than one process can operate simultaneously, depending which crystal field levels interact with phonons.  The time and temperature scales of this type of process mean they can be studied by neutron scattering.  Either the width ($\Gamma$) of the quasielastic neutron scattering (QENS) can be understood as lifetime broadening of the ground state doublet and used to give a measure of the spin relaxation time ($\tau$), as was done for rare earth cuprates~\cite{Lovesey:2000jh}; or the width of a CFE can be followed directly, as was done for LiTmF$_4$~\cite{Babkevich:2015wj}.  In the former case the origin of the relaxation was debated~\cite{Boothroyd:2001im}, while in the latter full details of the coupled phonons were not established.  In the following, we measure $\Gamma$ using QENS, determine the allowed spin-lattice interactions and construct a model of phonon mediated processes in both materials that describes the high temperature processes quantitatively.  We provide microscopic evidence of one such coupling.

\begin{figure}
   \includegraphics[trim=1 1 1 1,clip=true, scale=1.0]{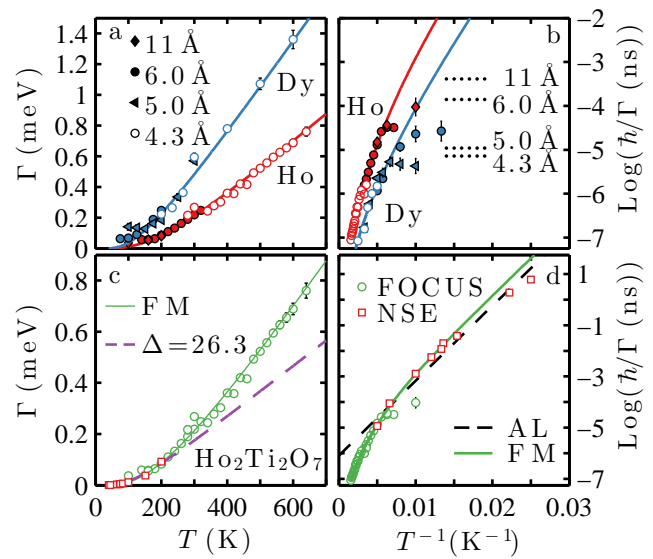}\\ 
  \caption{The QENS width $\Gamma$ as a function of temperature (a) and relaxation time as a function of inverse temperature (b), measured with different neutron wavelengths shown by the symbols.  Solid lines are from the  model described in the text, dotted lines in b indicate the resolution limit of the different settings of the spectrometer.  The same quantities for \hto{} are shown in panel c and d, incorporating QENS (FOCUS, this study) and  neutron spin echo (NSE,~\cite{Ehlers:2002jz}), compared to the full model (FM), the first term of the model ($\Delta=26.3$ meV), and an Arrhenius law (AL)~\cite{Ehlers:2002jz}.}
   \label{fig:f2}
\end{figure}

We have measured QENS in powder samples of both \hto{} and \dto{}~\cite{sup_mat} over a wide range of temperatures using the spectrometer FOCUS~\cite{Juranyi:2003jq} at SINQ.  We report results obtained using the $(0,0,2)$ reflection of both the pyrolytic graphite ($\lambda=4.3,5,6$ \AA; resolution $\approx100,50,40$ $\mu$eV) and mica ($\lambda=11$ \AA; resolution $\approx 20$ $\mu$eV) monochromators, where we selected the wavelength to give appropriate resolution for a range of temperatures.  The quasielastic scattering was fitted by a single Lorentzian, adjusted by the Bose factor for detailed balance.  The elastic line was removed by fitting with a Gaussian peak, whose parameters were derived from a measurement of the resolution using a vanadium standard.  Additional Lorentzians were incorporated in the fit of high temperature data from \hto{} to model excited state CFEs that appear close to the elastic line.  Points at the edges of two resolution ranges were measured with both settings to ensure overlap of the fitted peak widths.

\begin{figure*}
  \centering
\includegraphics[trim=1 1 1 1,clip=true, scale=1]{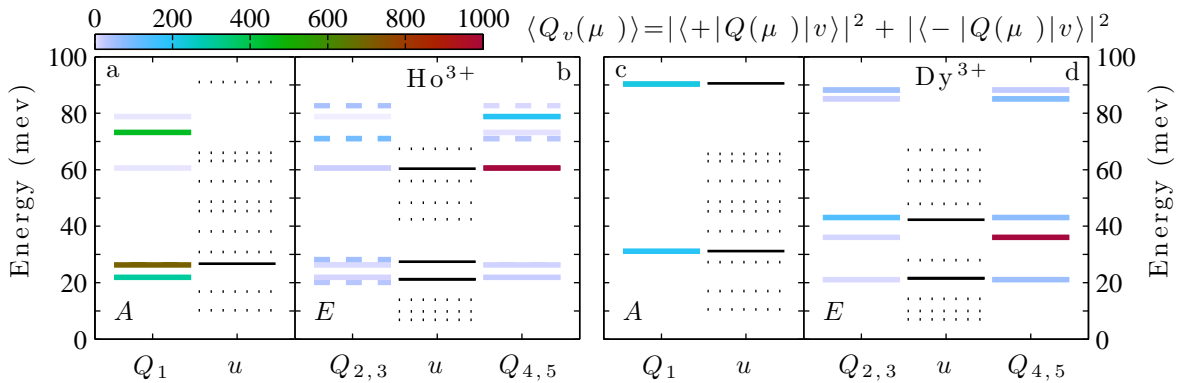}%
  \caption{Schematic overview of the interaction of CFEs and phonons. The line color of the CFEs represent the combined values of the quadrupolar transition matrix elements between the members of the crystal field ground state and excited states. Quadrupolar operators $Q_\mu$ and zone center phonon modes ($u$) of \hto{} (a and b) and \dto{} (c and d) are sorted by their symmetries $A$ (a and c) and $E$ (b and d). The relevant phonon modes (represented by solid lines) are quasi-degenerate with a CFE supporting a strong quadrupolar matrix element of the correct symmetry. Phonons that are not involved are shown by dotted lines.}
   \label{fig:f3}
\end{figure*}

An example of the $S(|\vec{Q}|,\omega)$ data obtained for \hto{} is shown in Fig.~\ref{fig:f1}a, and an example of the $|\vec{Q}|$-integrated data used for fitting is shown in Fig.~\ref{fig:f1}b.  The temperature dependence of the width and intensity of the quasielastic scattering and CFEs can be seen in Fig.~\ref{fig:f1}c.  Below $T\approx50$ K, the spin fluctuation processes are too slow for QENS, and, even with $\lambda=11$ \AA, the response is resolution-limited, but as the temperature is further increased, the QENS broadens.

In Fig.~\ref{fig:f2}a, we show the temperature evolution of $\Gamma$ for both compounds, and also its representation as a relaxation time $\tau$ (panel b).  Notably, the QENS spectra of \dto{} are nearly twice as broad as those of \hto{} throughout the sampled temperature range.  In Fig.~\ref{fig:f2}c,d we show our data for \hto{} compared with NSE data from Ref.~\cite{Ehlers:2002jz}, which extends to  longer times/lower temperatures, in terms of $\Gamma$ and $\tau$ respectively.  All the lines in Fig.~\ref{fig:f2} are derived from models, either the  model which we discuss below, or the Arrenhius law used in Ref.~\cite{Ehlers:2004ui}.   It can be seen in Ref.~\cite{Ehlers:2004ui} that the relaxation time already departs from the Arrhenius law at the highest temperatures studied there, and this is made plain by the higher temperatures measured in this work (see Fig.~\ref{fig:f2}c and d).

The phonon-mediated spin relaxation mechanism depends on a magnetoelastic interaction of normal modes of vibration with the single ion crystal field potential~\cite{Lovesey:2000jh}.  The contribution to the temperature dependence of $\Gamma$ is given by
\begin{equation}
\Gamma(T)=\sum_i\frac{3\pi r n_i}{2M\Delta_i}\zeta_{\mu}^2Z_\mu(\Delta_i)\{|\langle a |Q_\mu |v_i\rangle|^2+|\langle b |Q_\mu|v_i\rangle|^2\},
\label{eq:Orbacheq}
\end{equation}
where $\zeta_{\mu}$ is the magnetoelastic coupling parameter for a phonon and intermediate crystal field state $|v_i\rangle$ at energy $\Delta_i$, $r$ is the number of ions per unit cell, and $M$ the mass of an oxide ion.  The distribution function $n_i=(\exp{(\Delta_i/k_BT)}-1)^{-1}$ provides the temperature dependence of the process, and $Z_{\mu}(\Delta_i)$ is the partial phonon density of states (pPDOS) of the anionic modes of vibration transforming according to the representation $\mu$.  The prefactors are absorbed into fitting parameters such that $\Gamma(T)=\sum_{j}B_in_i$~\cite{Lovesey:2000jh}.  

For an intermediate state $i$ to enter the summation, we require a finite matrix element for the quadrupolar operator ($Q_\mu$) for the transition between the initial ($\langle a|$) and intermediate ($|v\rangle$) crystal field state and spectral overlap of this state with a phonon ($u_\mu$) of identical symmetry ($\mu$ labels the irreducible representation of the operator or excitation).  At the rare earth site, in $D_{3d}$ symmetry, there are three quadrupolar operators with symmetry $(A_1,E,E)$, and the matrix element for a transition $\langle a|Q_\mu|v\rangle$  is finite if the direct product of the representations ($\gamma$) of the two states and the transition operator contain the unit representation, $\gamma_a\times\gamma_v\times\gamma_Q\in A_1$.  There are three possible combinations for finite matrix elements of the magnetoelastic interaction operator: $\gamma_{a,Q,v,u}=E$ (1); $\gamma_{a,v}=E,\gamma_{Q,u}=A$ (2); $\gamma_{a,Q,u}=E,\gamma_v=A$ (3).  The initial state $|a\rangle$ is a member of the ground state doublet, and $|v\rangle$ is an excited crystal field state, the final state $|b\rangle$ is the other member of the ground state doublet.  

The transitions involved in the model are summarized in Fig.~\ref{fig:f3}.  Using the wave functions of crystal field states~\cite{xtal_fields}, we evaluated the quadrupolar matrix elements of the crystal field transitions.  Quasi-degenerate (at the Brillouin zone center) phonon modes of the correct symmetry were identified from the phonon band structure~\cite{phonons}.  The vibrational modes involved are dominated by oxide ions in the $48f$ position, so we approximate $Z_\mu(\Delta_i)$ by the pPDOS of this site.  

In \hto{}, we find the two largest matrix elements between the ground state and the doublets at $E=26.3, 60$ meV, and weaker matrix elements between the ground state and the singlet and doublet at $E=21, 22$ meV.  Each of these transitions is quasi-degenerate with a phonon of appropriate symmetry, while the remaining transitions have no overlap with a vibrational mode of $E$ or $A$ symmetry.     For \hto{}, we construct our model using three intermediate states at $\Delta_i=21.5,26.3,60$ meV, where the first represents the effect of the weak matrix elements.  In \dto{}, the intermediate states are those at $E=21,31, 43$ meV.  The state at 91 meV also meets the symmetry requirements but is outside the temperature window of this study.  Other states have large matrix elements, but no compatible phonon.

%

For \hto{} we included both QENS and NSE data in the fit, and since the states $B_1$ and $B_2$ have similar energies and nearly identical pPDOS~\cite{phonons}, we related their values by the ratio of their quadrupolar matrix elements.  The resulting coefficients are $B_{1,2,3}=0.018,0.2,0.79$ meV.  For \dto{}, to reduce the number of fitting parameters, the values of the parameters $B_2$ and $B_3$ were linearly related using the energies of their CFEs, unity for the ratio of the oxygen phonon density of states, and their quadrupolar transition matrix elements. We obtained $B_{1,2,3}=0.23, 0.49, 0.38$ meV.  As shown in Fig.~\ref{fig:f2}, the model fits the relaxation rates of both compounds very well.  For \hto{}, relaxation via the level at 26.3 meV describes the QENS width effectively up to $T\approx250$ K, and the third intermediate state at $E=60$ meV dominates at higher temperatures.

\begin{figure}
\includegraphics[trim=1 1 1 1,clip=true, scale=0.5]{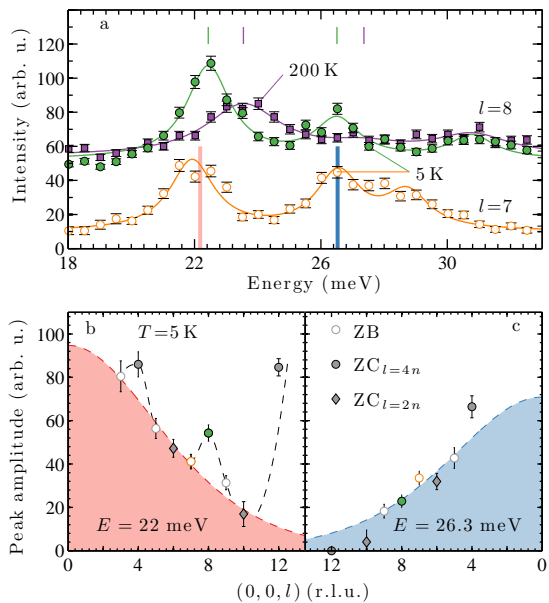}%
\caption{CFE and phonon interactions in a single crystal of \hto{}.  Two CFEs ($E=22,26.3$ meV) and a phonon ($E=31$ meV) can be seen.  The CFEs can be measured at the zone center ($l=8$) and boundary ($l=7$) and shift upward between 5 and 200 K.  The CFE at $E=22$ meV disperses upward by 0.5 meV at the zone center where it intersects with an optical phonon~\cite{phonons} (panel a).  The intensity of the two CFEs follows the magnetic form factor along $(0,0,l)$, except for the CFE at $E=22$ meV which is boosted at zone centers with strong phonon structure factors ($l=4n$) (panel b and c, scan positions of panel a are indicated by colored points).}
\label{fig:f5}
\end{figure}

The values for the magnetoelastic coupling constants $\zeta_{\mu}$ extracted from the fitted parameters under these approximations~\cite{sup_mat} suggest that the magnetoelastic coupling is linear (in energy), consistent with physical ingredients of the model.  
To further verify our model, we sought direct evidence of interactions between CFEs and phonons using a single crystal of \hto{} and the triple axis spectrometer HB3 at HFIR, ORNL.  The $(0,0,2)$ reflection of the beryllium monochromator provides access to quite high energy transfers with good energy resolution - using a pyrolytic graphite filter and analyzer ($(0,0,2)$ reflection), the energy resolution was $\Delta E \approx 1.7$ meV in the energy transfer window of $20-30$ meV.  With fixed final energies of $E_f=14.7, 30.5$ meV, we measured energy scans in the range $18<E<33$ meV at different $(0,0,l)$ positions that were either Brillouin zone centers $(0,0,l=2,4,6,8,10,12)$ or boundaries $(0,0,l=3,5,7,9)$, at $T=5,200$ K.

Fig.~\ref{fig:f5}a shows the two CFEs at $E\approx22,26.3$ meV and a phonon at $E\approx 31$ meV, measured at $(0,0,8)$.  The intensities of the CFEs decrease as the temperature is raised, and they shift upward in energy, while the intensity of the phonon increases but its energy does not change.  The upward shift of the CFEs is also shown by the downward shift of the excited state transitions T1 and T2 in Fig.~\ref{fig:f1}c.  Comparison of the same scan at $l=7,8$ shows a resolution limited sharp peak for both CFEs with a 0.5 meV upward dispersion between zone boundary and center for the first, but at identical energies for the second.  The $l$-dependence of the intensity of the CFEs (Fig.~\ref{fig:f5}b) follows the dipole magnetic form factor at zone boundaries ($l=n$) and at zone centers where the $Fd\bar{3}m$ space group forbids a Bragg reflection ($l=2n$), but the CFE at $E\approx22$ meV has anomalously large intensity at zone centers with strong Bragg reflections ($l=4n$) while the CFE at $E\approx 26.3$ meV also follows the magnetic form factor at these positions.  

Phonon calculations~\cite{phonons} show that there is an optical phonon with $E$ symmetry at $E\approx22$ meV at the zone center.  The phonon disperses away at the zone boundary, and its structure factor is suppressed at zone centers where the Bragg intensity is not allowed.  Hence at all these positions (i.e. $l=n$ and $l=2n$) both CFEs are unaffected and follow the magnetic form factor.  At those zone centers with a strong Bragg reflection, the strong phonon structure factor boosts the intensity well above the magnetic form factor, but the observation of a single mode displaced from the energy of the uncoupled zone boundary CFE or phonon shows that the coupling pulls the two excitations into resonance, i.e. they are not just coincident.  Conversely, the phonon mode expected to interact with the CFE at 26.3 meV was calculated to have a very weak structure factor along $(0,0,l)$, due to its polarization.  Hence we observe no signatures of coupling in this direction, and this CFE also follows the magnetic form factor (Fig.~\ref{fig:f5}c). 

We have shown that the symmetries and wavefunctions CFEs and optical phonons can be used to construct a physically realistic model for phonon mediated spin flipping processes.  Modes with the correct symmetry and energy exist in the spin ices \hto{} and \dto{}, and we presented direct evidence of one of the couplings in \hto{}.  A model based on these spin-lattice interactions describes the high temperature spin relaxation in both compounds very well.  We advance this model as the first microscopic description of a spin flipping mechanism in the spin ices \hto{} and \dto{}, and also as a quantification of the spin-lattice interactions possible in these materials.  Our investigation sets the stage for microscopic investigations of the possible quantum processes at low temperature, and their consequences for collective monopole dynamics.

\begin{acknowledgments}
MR and TF thank ORNL staff for support; P. Santini, B. Tomasello, C. Castelnovo, R. Moessner, J. Quintanilla, and S. Giblin for discussion; and the authors of Refs.~\cite{xtal_fields} and ~\cite{phonons} for related collaboration.  Neutron scattering experiments were carried out at the continuous spallation neutron source SINQ at the Paul Scherrer Institut at Villigen PSI in Switzerland; and High Flux Isotope reactor (HFIR) of Oak Ridge National Laboratory, Oak Ridge, Tennessee, USA.  Work at PSI was partly funded by the Swiss NSF (grants 200021\_140862 and 200020\_162626).  Research at Oak Ridge National Laboratory's HFIR was sponsored by the Scientific User Facilities Division, Office of Basic Energy Sciences, U. S. Department of Energy.
\end{acknowledgments}


%

\section{Supplementary Information}

{\it Further details of samples and experiments:} The powder samples of \hto{} and \dto{} were originally described in Refs.~\cite{phonons,xtal_fields}.  They were prepared from stochiometric ratios of the oxides Ho$_2$O$_3$ or Dy$_2$O$_3$, and TiO$_2$ in a solid state reaction.   The oxides, with 99.99\% purity, were annealed at 850 $^\mathrm{o}$C for 10 hours, then mixed and ground, and heated at 950-1300 $^\mathrm{o}$C for 140 hours with several intermediate grindings.  The structures were verified by combined neutron and x-ray diffraction experiments, which were carried out on HRPT~\cite{Fischer:2000to} at SINQ, PSI and the Materials Science Beamline (MSB)~\cite{Willmott:2013gv} at the SLS, PSI.    Rietveld refinement of the structures as implemented in the Fullprof~\cite{fullprof} software proved both samples to be of high quality and single phase.  

For the QENS experiment on FOCUS, to overcome the strong absorption of natural isotopic abundance dysprosium we used a flat plate sample holder with very thin layer of sample.  The plate was oriented perpendicular to the incident beam to reduce the path length of neutrons in the sample, and we used data from low angle detectors only.  To have reasonable statistics requires long counting times.  We also confirmed some points by measuring our isotopically enriched crystal of $^{162}$\dto{}, but the crystal is much smaller than the cross section of the beam so intersects relatively few incident neutrons and hence does not greatly improve the situation.  \hto{} can be used directly as a packed powder and the count rate is much larger.

\begin{figure}
\includegraphics[trim=1 1 1 1,clip=true, scale=0.8]{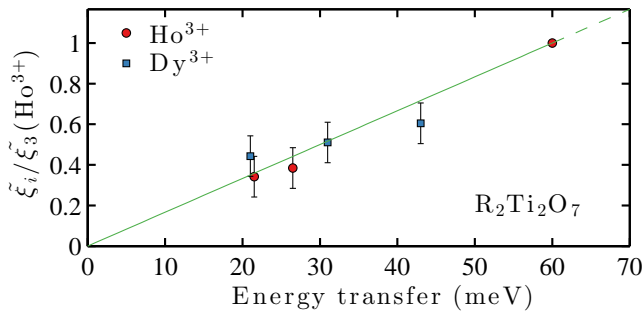}%
\caption{Magneotelastic coupling constants extracted from the fitted parameters $B_j$, using the partial phonon density of states for the O($48f$) for $Z_\mu(\Delta_j)$, normalizing the rare earth ions by the Stevens factors $\alpha(J)$, and scaling to highest energy coupling in \hto{}.}
\label{fig:s1}
\end{figure}

{\it Magnetoelastic coupling constants:} The determination of an absolute value for the magnetoelastic coupling constant $\zeta$ from the fitted $B$ parameters is difficult. Despite having the full calculated pyrochlore phonon spectrum~\cite{phonons}, there is a large uncertainty connected with the evaluation of the partial phonon densities of states $Z_{\mu}(\Delta_i)$, which only includes individual vibrational modes that transform according to the representation $\mu$. We note however, that all vibrational modes involved in the model in both compounds are dominated by the motion of the oxygen O(48$f$) ions, which surround the rare earth ion. Hence, the partial phonon density of states of the  O(48$f$) ions gives at least good estimates for the ratios of the $Z_{\mu}(\Delta_i)$ at different energy transfers $\Delta_i$. We therefore estimate all the magnetoelastic coupling constants $\zeta_i$ for the intermediate CEF states in both \hto{} and \dto{} relative to the coupling constant $\zeta_3$ of the third intermediate CEF state in \hto{} at $E=60$\,meV:
\begin{equation}
\frac{\zeta_i}{\zeta_3} = \sqrt{\frac{B_i\Delta_i Z_{\text{O}(48f)}(\Delta_3) \left[ \sum_{\mu} \sum_{j=\pm}|\langle j|Q(\mu)|v_3^j\rangle|^2 \right]}{ B_3\Delta_3 Z_{\text{O}(48f)}(\Delta_i) \left[ \sum_{\mu} \sum_{j=\pm}|\langle j|Q(\mu)|v_i^j\rangle|^2 \right] }},
\end{equation}
where the summation $j$ runs over the two members of the corresponding ground state doublet and all members of the excited intermediate CEF state $v_i$. The ratios of the partial phonon densities of states are taken as: 
\begin{equation}
\frac{Z_{\text{O}(48f)}(\Delta=21\,\text{meV})}{Z_{\text{O}(48f)}(\Delta=60\,\text{meV})} = \frac{Z_{\text{O}(48f)}(\Delta=26\,\text{meV})}{Z_{\text{O}(48f)}(\Delta=60\,\text{meV})} = 1.5 \nonumber 
\end{equation}
\begin{equation}
\frac{Z_{\text{O}(48f)}(\Delta=31\,\text{meV})}{Z_{\text{O}(48f)}(\Delta=60\,\text{meV})} = 1.
\end{equation}
In order to compare between two different ions, we calculate the ratios of the bare magnetoelastic coupling constants $\tilde{\zeta_i}$, which follow from  $\zeta=\tilde{\zeta}\alpha{J}$ \cite{Lovesey:2000jh} with the Stevens factors $\alpha(J)$ of the rare earth ions, here $\alpha=-1/450$ and $-2/315$ for Ho$^{3+}$ and Dy$^{3+}$ respectively. The resulting ratios are illustrated in Fig.~\ref{fig:s1}. We find that the bare magnetoelastic coupling constants of all intermediate CEF states in the two compounds depend linearly on the energy transfer of the intermediate states within errorbars. Given a linear mode coupling such as between the quadrupolar operators and the normal modes of vibration in the magnetoelastic interaction operator of Eq. \ref{eq:Orbacheq}, it is plausible that the magnetoelastic coupling strength vanishes in the limit of zero energy transfer and is generally highest at high energy transfer. The experimental observations are therefore consistent with the expectation and give strong support for the phonon mediated mechanism with the established intermediate CEF states in both rare earth pyrochlore titanates \hto{} and \dto{}.

{\it Relation with previously described relaxation processes:} Previously two relaxation processes have been reported in both compounds~\cite{Snyder:2001uu,Matsuhira:2001cp,Snyder:2003ek,Snyder:2004it,Ehlers:2002jz}.  In \dto{} these can be seen in $\chi_{ac}$ experiments because the relative rates of them are favorable for this technique.  The low temperature process, ascribed to quantum tunneling due to the plateau in the relaxation time at intermediate temperatures, is sufficiently slow that it does not compete with the high temperature process even as this becomes slow at temperatures of $T\approx20$ K.  Hence peaks can be observed in $\chi''$ (as measured by $\chi_{ac}$) as each process dominates the relaxation in the experimental frequency range, first the high temperature process at $T\approx20$ K, then the low temperature process at $T<10$ K.  In \hto{}, the low temperature process seems to be relatively much faster, and competes with the high temperature process effectively on the time scales of $\chi_{ac}$.  Therefore, one cannot observe a peak in $\chi''$ that corresponds to the high temperature process using $\chi_{ac}$ measurements.  To observe the high temperature process one must go to higher temperatures where the low temperature process is no longer competitive and the high temperature process is necessarily much faster.  This means that techniques with shorter timescales are required to observe it, which is why it was originally found using neutron spin echo and quasielastic neutron scattering~\cite{Ehlers:2002jz,Ehlers:2004ui}.

The high temprature process in \dto{} was fitted by an Arrhenius law with activation energy $\Delta=17-25$ meV, which was compared to the first CFE.  At this time, the exact energy of the CFE was not actually known, but was expected to be of this order.  It is now known to be at $21$ meV~\cite{xtal_fields}.  In \hto{}, the activation energy was given as $\Delta=25(\pm1)$ meV, but the CFEs are at $E=21.9, 26.3$ meV.  The high temperature process was described as a flipping of the spin via the first CFE, but in this case the activation energy should match exactly the energy of the CFE, and the Arrhenius law should describe the relaxation time throughout the temperature range where the process is dominant.  Neither of these conditions is actually fulfilled, particularly when the relaxation time is measured to higher temperatures.  

However, the previous investigations showed that the presence or absence of a peak in $\chi''_{ac}$ depends on the relative rate of high and low temperature processes.  In absence of a competing low temperature process \hto{} would also have a peak in $\chi''_{ac}$ for the high temperature process at a similar temperature to \dto{}.  This condition is still satisfied if the temperature dependence of the relaxation rate provided by our model is treated in an analogous fashion.  Hence the phonon-mediated model provides a physical explanation and quantitative description of all known observations about the high temperature processes.

\end{document}